\documentclass[article, reprint, superscriptaddress, amsmath, amssymb]{revtex4-2}

\usepackage[utf8]{inputenc}
\usepackage[T1]{fontenc}
\usepackage{newtxtext,newtxmath} 

\usepackage{graphicx}   
\usepackage{dcolumn}    
\usepackage{bm}         
\usepackage{tabularx}   
\usepackage{siunitx}    
\usepackage{float}

\usepackage{xcolor}     
\usepackage{soul}       
\usepackage{textcomp}   

\usepackage{comment}

\usepackage[mathlines]{lineno} 

\graphicspath{{}}

\newcommand{\ba}{\begin{eqnarray}}
\newcommand{\ea}{\end{eqnarray}}

\newcommand{\NbSe}{$\text{NbSe}_2$}
\newcommand{\WSe}{$\text{WSe}_2$}

\begin{document}

\title{Coherent and compact van der Waals transmon qubits}
\author{Jesse Balgley}
\altaffiliation{These authors contributed equally to this work.}
\affiliation{Department of Mechanical Engineering, Columbia University, New York, NY 10027, USA}
\author{Jinho Park}
\altaffiliation{These authors contributed equally to this work.}
\affiliation{Department of Mechanical Engineering, Columbia University, New York, NY 10027, USA}
\author{Xuanjing Chu}
\altaffiliation{These authors contributed equally to this work.}
\affiliation{Department of Applied Physics and Applied Mathematics, Columbia University, New York, NY 10027, USA}
\author{Jiru Liu}
\affiliation{Department of Physics and Astronomy, Northwestern University, Evanston, IL 60208, USA}
\author{Madisen Holbrook}
\affiliation{Department of Physics, Columbia University, New York, NY 10027, USA}
\author{Kenji Watanabe}
\affiliation{Research Center for Electronic and Optical Materials, National Institute for Materials Science, 1-1 Namiki, Tsukuba 305-0044, Japan}
\author{Takashi Taniguchi}
\affiliation{Research Center for Materials Nanoarchitectonics, National Institute for Materials Science,  1-1 Namiki, Tsukuba 305-0044, Japan}
\author{Archana Kamal}
\affiliation{Department of Physics and Astronomy, Northwestern University, Evanston, IL 60208, USA}
\author{Leonardo Ranzani}
\affiliation{RTX BBN Technologies, Quantum, Photonics, and Computing Group, Cambridge, MA 02138, USA}
\author{Martin V.~Gustafsson}
\affiliation{RTX BBN Technologies, Quantum, Photonics, and Computing Group, Cambridge, MA 02138, USA}
\author{James Hone}
\affiliation{Department of Mechanical Engineering, Columbia University, New York, NY 10027, USA}
\author{Kin Chung Fong}
\affiliation{RTX BBN Technologies, Quantum, Photonics, and Computing Group, Cambridge, MA 02138, USA}
\affiliation{Quantum Materials and Sensing Institute, Northeastern University, Burlington, MA 01803, USA}
\affiliation{Department of Electrical and Computer Engineering, Northeastern University, Boston, MA 02115, USA}
\affiliation{Department of Physics, Northeastern University, Boston, MA 02115, USA}
\date{\today}

\begin{abstract}
State-of-the-art superconducting qubits rely on a limited set of thin-film materials. Expanding their materials palette can improve performance, extend operating regimes, and introduce new functionalities, but conventional thin-film fabrication hinders systematic exploration of new material combinations. Van der Waals (vdW) materials offer a highly modular crystalline platform that facilitates such exploration while enabling gate-tunability, higher-temperature operation, and compact qubit geometries. Yet it remains unknown whether a fully vdW superconducting qubit can support quantum coherence and what mechanisms dominate loss at both low and elevated temperatures in such a device. Here we demonstrate quantum-coherent merged-element transmons made entirely from vdW Josephson junctions. These first-generation, fully crystalline qubits achieve microsecond lifetimes in an ultra-compact footprint without external shunt capacitors. Energy relaxation measurements, together with microwave characterization of vdW capacitors, point to dielectric loss as the dominant relaxation channel up to hundreds of millikelvin. These results establish vdW materials as a viable platform for compact superconducting quantum devices.

\end{abstract}

\maketitle




Superconducting qubits have become a leading platform for realizing fault-tolerant quantum computing, owing to a dramatic improvement in qubit lifetimes by roughly six orders of magnitude since their earliest demonstrations~\cite{kjaergaard_superconducting_2020}. This progress was driven by a deeper understanding of loss mechanisms within constituent materials~\cite{deleon_materials_2021,murray_material_2021}, along with design strategies to suppress loss channels. Yet the materials used in state-of-the-art superconducting qubits have remained largely unchanged --- most still rely on amorphous aluminum and its oxide. The small superconducting gap and $1.2~\text{K}$ critical temperature of aluminum increases susceptibility to quasiparticle poisoning~\cite{wang_measurement_2014}, while disordered materials and interfaces introduce dielectric loss and device variability through two-level-system defects~\cite{bayros_influence_2024}. Importantly, these material constraints limit superconducting qubit operating temperatures, demanding cryogenic overhead that may be impractical at scale.

Expanding the materials palette for superconducting qubits offers opportunities not only to mitigate disorder and improve scalability, but also to unlock new device functionalities. Semiconductor weak links, for example, support magnetic-field-compatible qubits~\cite{kringhoj_magnetic-field-compatible_2021}, voltage tunability~\cite{larsen_semiconductor_2015,casparis_superconducting_2018,pita-vidal_gate-tunable_2020,strickland_gatemonium_2025}, Andreev-bound-state and hybrid charge–spin physics~\cite{janvier_coherent_2015,hays_coherent_2021}, and protected qubits~\cite{larsen_parity-protected_2020}. Superconductors with higher critical temperatures extend operation to higher frequencies and temperatures~\cite{anferov_superconducting_2024,anferov_millimeter_2025,wang_all-nitride_2025_arx}, while high-kinetic-inductance materials provide the reactances needed for fluxonium and $0$--$\pi$ qubits~\cite{hazard_nanowire_2019,grunhaupt_granular_2019}. Ultimately, introducing new materials into superconducting qubits will require demonstrating that they can support robust coherence, motivating systematic testing of emerging materials and material combinations.

Van der Waals (vdW) materials present a promising, highly modular route toward this goal. The broad range of superconductors, semiconductors, semimetals, and insulators spanned by vdW materials can be obtained as single crystals with low defect densities~\cite{liu_two-step_2023}. Moreover, they can be assembled through layer-by-layer stacking~\cite{wang_one-dimensional_2013} to form heterostructures with atomically pristine interfaces independent of lattice matching~\cite{dean_hofstadter_2013}, allowing for a uniquely modular design of devices with arbitrary material combinations. These attributes have enabled vdW Josephson weak links for gate-tunable qubits~\cite{wang_coherent_2019}, resonators~\cite{schmidt_ballistic_2018}, parametric amplifiers~\cite{sarkar_quantum-noise-limited_2022,butseraen_gate-tunable_2022}, and single-photon detectors~\cite{walsh_graphene-based_2017}. More recently, vdW parallel-plate capacitors have reduced transmon footprints while leveraging superconductors with critical temperatures six times higher than aluminum~\cite{antony_miniaturizing_2021,wang_hexagonal_2022}. Vertical vdW superconductor--semiconductor Josephson junctions allow robust, uniform tunnel barriers roughly ten times thicker than aluminum oxide~\cite{balgley_crystalline_2025}. These advances guide the realization of a fully crystalline qubit in which the Josephson junction itself provides both the nonlinearity and the capacitance of a transmon, an architecture known as the merged-element transmon (MET)~\cite{zhao_merged-element_2020}. In this compact qubit, electromagnetic fields are confined to the small mode volume of a vertical junction, emphasizing the need for high-quality ordered materials and interfaces. Whether vdW materials can meet these requirements, support quantum coherence, and push superconducting qubits to higher-temperature operation remain open questions.

In this work, we design, fabricate, and measure \emph{all-vdW} merged-element transmons and investigate the loss mechanisms within these devices. Through careful material selection and circuit design, we demonstrate coherence in fully crystalline qubits and identify dielectric loss as the dominant relaxation channel, even at elevated temperatures. Together, these results establish vdW superconductor--semiconductor vertical Josephson junctions as a viable platform for compact, versatile superconducting qubits.

\begin{figure*}[ht]
\centering
\includegraphics[width=2\columnwidth]{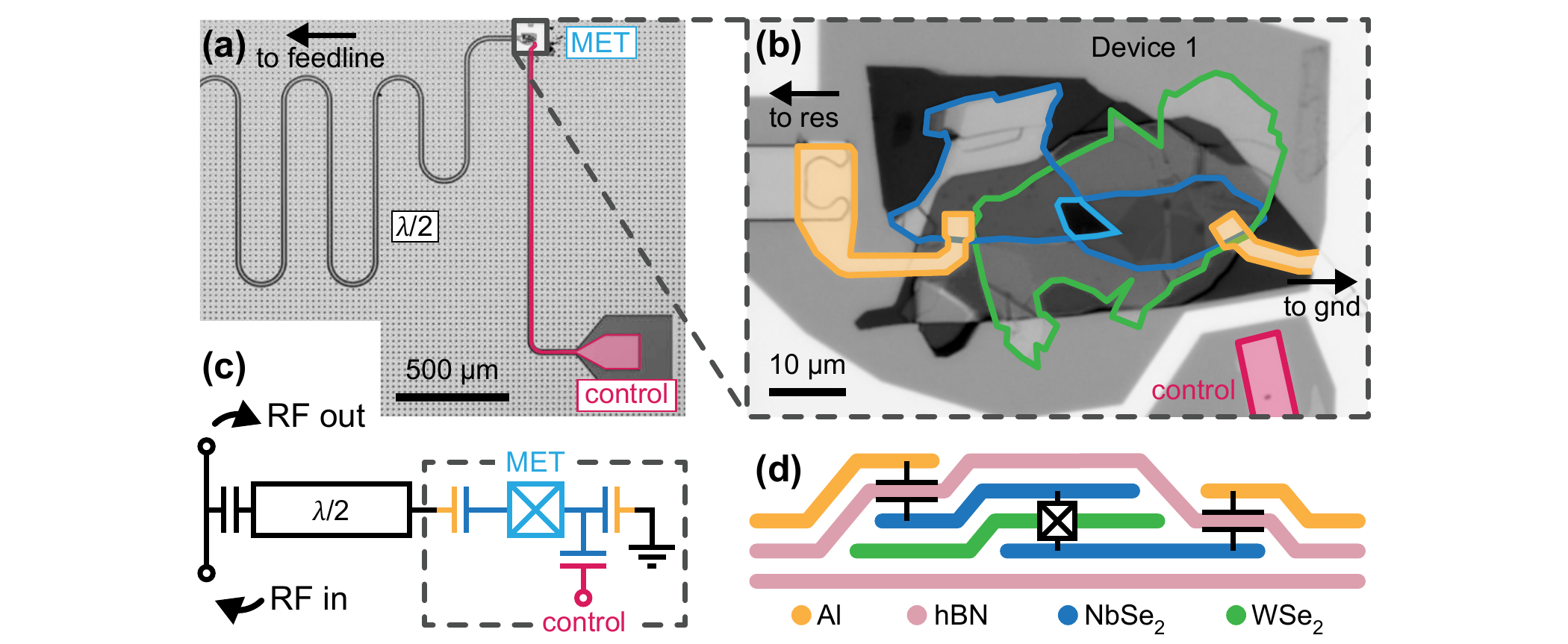}
\caption{Van der Waals merged-element transmon. (a), (b) Optical micrographs of Device 1. A Josephson junction (light blue) is formed by the overlap of two \NbSe\ flakes (dark blue) sandwiching a flake of \WSe\ (green). The device is encapsulated from top and bottom by flakes of hBN, which are not outlined. Aluminum wires (orange) capacitively couple the MET to a $\lambda/2$ CPW readout resonator (res) and ground (gnd). A nearby wire is used to send control pulses (red). The readout resonator is capacitively coupled to a CPW feedline outside of the field of view. (c) Circuit diagram of the qubit and readout circuit. RF signals are sent through the feedline to measure the complex transmission coefficient. (d) Side-view schematic of a floating vdW MET, including the vdW Josephson junction stack and Al coupling electrodes (not to scale).}
\label{fig:Device}
\end{figure*}

\section*{Results}
\subsection*{Qubit Design and Fabrication}
To design a superconducting qubit from a new materials platform, we must target circuit parameters that satisfy the requirements of an MET. In this architecture, the vdW Josephson junction provides both the nonlinearity and capacitance to achieve an anharmonic energy spectrum. In a transmon, generally, the charging energy \mbox{$E_C = e^2/2C_\text{total}$} and Josephson energy \mbox{$E_J = \Phi_0 I_c / 2\pi$} determine the qubit frequency \mbox{$f_{01} = \left(\sqrt{8E_JE_C} + \alpha\right)/h$} and anharmonicity \mbox{$\alpha = -E_C$}~\cite{koch_charge-insensitive_2007}, where $I_c$ is the junction critical current and $C_\text{total}$ is the total qubit capacitance. In an MET, the the junction capacitance $C_J$ dominates \mbox{$C_\text{total} \approx C_J + C_g$}, with $C_g$ representing coupling to the readout circuitry and to ground. Because of the compact vertical geometry of the MET, we assume stray capacitances are negligible compared to $C_J$ and $C_g$.

Our METs are realized by mechanically stacking vdW materials to form vertical Josephson junctions. The junction electrodes are 30--40-nm-thick flakes of the vdW superconductor \NbSe, which has a critical temperature $T_c \approx 7~\text{K}$ (Ref.~\cite{khestanova_unusual_2018}). The tunnel barrier is \WSe, a vdW semiconductor with a $1.2~\text{eV}$ bandgap. Crucially, since the Josephson coupling is governed by quantum tunneling across the barrier, this relatively small bandgap permits tunnel barriers approximately ten times thicker than those required for large-bandgap insulators such as aluminum oxide or vdW hexagonal boron nitride (hBN) to achieve the desired $I_c$~\cite{balgley_crystalline_2025}. The lateral overlap of the \NbSe\ electrodes defines the junction area $A$, which can be set during stacking using motorized micromanipulators or adjusted afterward by subtractive etching. The barrier thickness $t$ is controlled by selecting \WSe\ flakes of known layer number, with each atomic layer approximately $6.5~\text{\AA}$ thick~\cite{hicks_semiconducting_1964}. Each device is encapsulated in hBN to prevent oxidation and contamination during subsequent processing. Additional fabrication details are provided in Appendix~\ref{app:methods}.

Targeting specific qubit parameters requires careful calibration of the critical current as a function of $t$. From DC measurements of twenty \NbSe/\WSe\ Josephson junctions, we previously characterized this dependence over six orders of magnitude in $I_c$ for \WSe\ weak links between 2 and 12 nm thick~\cite{balgley_crystalline_2025}. Using the empirical $I_c$--$t$ relation, we calculate the expected $f_{01}$ of the vdW junctions as a function of $t$ by assuming a parallel-plate junction capacitance \mbox{$C_J = \kappa \epsilon_0 A / t$}, where $\kappa = 7.8$ (Ref.~\cite{laturia_dielectric_2018}) and $\epsilon_0$ the vacuum permittivity. This analysis indicates that a 17-layer-thick (\mbox{$t \approx 11~\text{nm}$}) \WSe\ barrier should yield the desired $f_{01} \approx 6~\text{GHz}$ without requiring an external shunt capacitance. For this barrier thickness, junction areas of $10$--$30~\text{\textmu m}^2$ provide sufficient capacitance to achieve anharmonicities $\alpha/h = -100$ to $-300~\text{MHz}$, and $E_J/E_C$ ratios between 50 and 100, ensuring insensitivity to charge noise~\cite{koch_charge-insensitive_2007}. Because each vdW MET is individually assembled, fabricated devices may exhibit deviations from designed values due to stacking tolerances and uncertainties in junction geometry (see Appendix~\ref{app:methods}).


\begin{figure*}[t!]
\centering
\includegraphics[width=2\columnwidth]{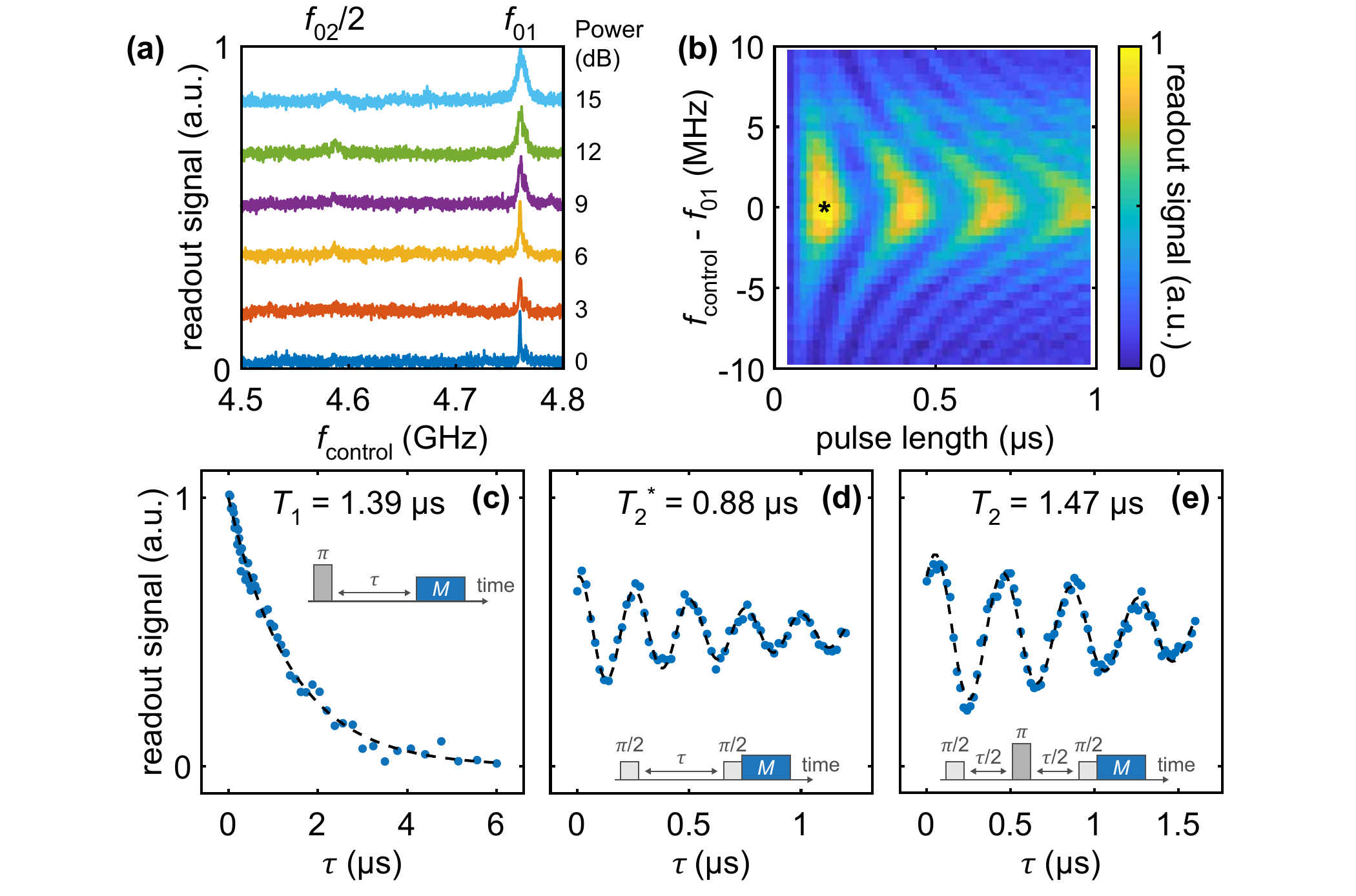}
\caption{MET spectroscopy and coherence in Device~1. (a) Two-tone spectroscopy showing the $0\rightarrow1$ qubit transition at $f_{01}$ and two-photon $0\rightarrow2$ transition at $f_{02}/2$. (b) Rabi oscillations around $f_{01}$. The black asterisk indicates the pulse length for a $\pi$ pulse. (c)--(e), Best-performing population inversion, Ramsey, and Hahn echo measurements, respectively. Black dashed lines are fits to the data, assuming exponential decay, from which qubit lifetimes are extracted (see main text). Schematic pulse sequences, including $\pi$ (dark grey), $\pi/2$ (light grey), and measurement $M$ (blue) pulses, are inset in each plot.}
\label{fig:Coherence}
\end{figure*}

Figure~\ref{fig:Device}(a) depicts the vdW MET positioned on a silicon substrate (dark gray) and integrated into a niobium circuit layer (light gray) that forms both the $\lambda/2$ coplanar-waveguide (CPW) readout resonator and its ground plane. A dedicated control line (red) capacitively delivers microwave drive pulses to the qubit. A magnified view in Fig.~\ref{fig:Device}(b) highlights the vdW material stack, coupling electrodes, and control wiring that define the MET. The full measurement architecture is shown in Fig.~\ref{fig:Device}(c), where radio-frequency (RF) signals propagate along a CPW feedline coupled to the readout resonator. Figure~\ref{fig:Device}(d) illustrates a schematic cross-section of the vdW junction and its coupling electrodes. Qubit coupling is implemented using parallel-plate capacitors formed between deposited metal leads and the \NbSe\ junction electrodes, with the top encapsulating hBN flake serving as the dielectric. This vertical coupling confines electromagnetic fields to compact, well-defined regions, minimizing stray capacitance. In most transmon architectures, including all previous MET demonstrations~\cite{zhao_merged-element_2020, mamin_merged-element_2021, daum_investigation_2025_arx}, qubit coupling is achieved laterally through coplanar capacitance patterned in the same plane as the qubit electrodes. This geometry is favored with conventional materials because vertical coupling through amorphous oxide dielectrics introduces additional loss, yet it typically increases the individual qubit footprint. In contrast, the use of a crystalline vdW dielectric enables low-loss vertical capacitive coupling~\cite{antony_miniaturizing_2021,wang_hexagonal_2022}. Beyond improving compactness, this approach offers key advantages for scalability: the localized vertical field profile should significantly reduce stray capacitances and interqubit crosstalk compared to extended lateral coupling designs. We design $C_g$ such that $\chi/h$ is approximately half the resonator linewidth~\cite{krantz_quantum_2019} ($\sim600~\text{kHz}$ here). The resulting junction energy participation ratios, \mbox{$p = C_J / C_\text{total}$}, exceed 90\% for all METs studied in this work (see Table~\ref{tab:qubit_properties}), ensuring that fields are predominantly concentrated in the vdW junction.


\begin{table*}[tb]
\caption{\label{tab:qubit_properties}Properties of van der Waals merged-element transmons.}
\begin{ruledtabular}
\begin{tabular}{cccccccccccccccc}
 & Area & $C_g$ & $p$ & $f_{01}$ & $\alpha/h$ &  
  & \multicolumn{3}{c}{$T_1$ ({\textmu}s)} 
  & \multicolumn{3}{c}{$T_2^*$ ({\textmu}s)} 
  & \multicolumn{3}{c}{$T_{2E}$ ({\textmu}s)} \\
\cline{8-10} \cline{11-13} \cline{14-16}
Device & ({\textmu}m$^2$) & (fF) & (\%) & (GHz) & (MHz) & $E_J/E_C$ 
      & Mean & Std.~dev. & Best 
      & Mean & Std.~dev. & Best 
      & Mean & Std.~dev. & Best \\
\hline
1 & 20 & 4  & 97 & 4.76 & $-174$ & 101 &
1.11 & 0.12 & 1.39 &
0.70 & 0.07 & 0.88 &
1.26 & 0.09 & 1.47 \\
2 & 8  & 2.4 & 96 & 5.79 & $-618$ & 14 &
1.67 & 0.27 & 2.46 &
0.92 & 0.53 & 2.77 &
3.29 & 0.74 & 4.82 \\
3 & 24 & 12  & 91 & 5.29 & $-100$ & 45 &
0.27 & 0.05 & 0.38 &
0.40 & 0.13 &      &
0.29 & 0.10 &      \\
4 & 20 & 3    & 98 & 5.15 & $-242$ & 66 &
&      & 0.20 &
&      & 0.16 &
&      &      \\
\end{tabular}
\end{ruledtabular}
\end{table*}

\subsection*{Qubit Coherence}

We present results for four coherent vdW METs, labeled Devices~1--4. These qubits span a wide range of $E_J/E_C$ ratios, anharmonicities, circuit configurations, and fabrication procedures, enabling a systematic study of how design parameters and device geometry influence coherence. Devices~1, 2, and 4 are electrically floating, with the bottom \NbSe\ electrode capacitively coupled to ground, whereas Device~3 is grounded on one side through a superconducting connection between the bottom \NbSe\ electrode and the aluminum wiring layer. Devices~1--3 are fixed-frequency (single-junction) transmons, while Device~4 --- for which fabrication details and spectroscopic characteristics were previously reported in Ref.~\cite{balgley_crystalline_2025} --- is patterned into two equal-area junctions to form a superconducting quantum interference device (SQUID) loop, allowing magnetic-flux tunability $f_{01}$ (see Appendix \ref{app:methods}). A nearby aluminum bias line provides DC flux control, and all measurements reported for this device are performed at the qubit’s flux ``sweet spot'', where $f_{01}$ is insensitive to fluctuations in magnetic flux. Additionally, the chips for Devices~2 and 3 were given acid treatment prior to placing the vdW stack (see Appendix \ref{app:methods} for details), whereas the other devices were fabricated on chips cleaned only with solvents. Measurements of Devices~1--4 thus allow us to gain insight into the influence of device configuration and fabrication conditions on qubit coherence.

In Fig.~\ref{fig:Coherence}, we show representative measurements of Device~1 at a temperature \mbox{$T = 15~\text{mK}$}. Figure~\ref{fig:Coherence}(a) depicts two-tone spectroscopy, in which a pulse of varying frequency is applied via the control line, followed by a readout pulse sent through the feedline to probe the frequency shift of the resonator near $f_{RO}$. 
At low control power (darker blue trace), a single peak in the readout signal reveals the $0\rightarrow1$ qubit transition at frequency $f_{01} \approx 4.76~\text{GHz}$. With increasing control power, the $f_{01}$ peak broadens, and a second peak appears at frequency $f_{02}/2\approx4.59~\text{GHz}$, corresponding to the two-photon $0\rightarrow2$ transition. From these transition frequencies, we extract the qubit anharmonicity (and equivalently the charging energy) as \mbox{$\alpha/h = f_{02} - 2f_{01} = -174~\text{MHz}$} and the Josephson energy $E_J$. The resulting \mbox{$E_J/E_C = 101$} places this qubit well within the transmon regime. Table \ref{tab:qubit_properties} lists measured qubit parameters for Devices~1--4.

Next, we measure Rabi oscillations (Fig.~\ref{fig:Coherence}(b)) by applying a control pulse of varying length and frequencies followed by a readout pulse. The resulting chevron pattern confirms the presence of a coherent two-level system centered at $f_{01}$ and provides the calibration parameters needed to generate a $\pi$ pulse for subsequent lifetime measurements.

Figures~\ref{fig:Coherence}(c)--(e) show representative population-inversion, Ramsey, and Hahn-echo measurements, respectively, for Device~1. From statistics over $100$ repetitions (see Appendix~\ref{app:methods}), we extract mean lifetimes of $T_1 = 1.11 \pm 0.12~\text{\textmu s}$, $T_2^* = 0.70 \pm 0.07~\text{\textmu s}$, and $T_2 = 1.26 \pm 0.09~\text{\textmu s}$, where associated uncertainties represent the standard deviations of the $T_1$ distributions. Across Devices~1--3, we find $T_2 \gtrsim T_1$, indicating that energy relaxation is the primary decoherence mechanism. Lifetimes for all four qubits are summarized in Table~\ref{tab:qubit_properties}. Interestingly, Device~2 exhibits the longest lifetimes ($T_1 = 1.67 \pm 0.27~\text{\textmu s}$, $T_2^* = 0.92 \pm 0.53~\text{\textmu s}$, and $T_2 = 3.29 \pm 0.74~\text{\textmu s}$) despite its comparatively low $E_J/E_C = 14$, which should make it more susceptible to charge fluctuations~\cite{schreier_suppressing_2008}.
\begin{figure*}[t!]
\centering
\includegraphics[width=2\columnwidth]{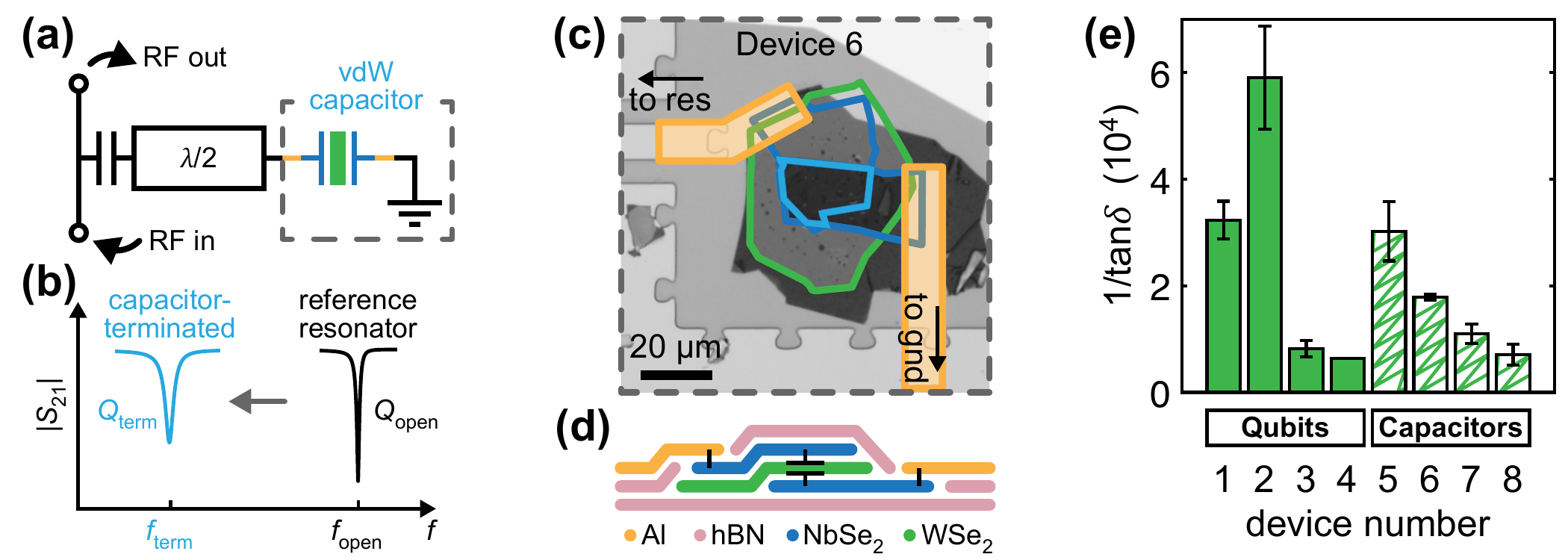}
\caption{Dielectric loss in van der Waals capacitors. (a) Circuit diagram of a CPW resonator (black rectangle) loaded with a vdW parallel plate capacitor consisting of \NbSe\ electrodes (dark blue) and \WSe\ dielectric (green). (b) Example of how a capacitive termination affects the resonance of a CPW resonator. (c) Optical micrograph of a vdW capacitor electrically connected to the resonator and ground via aluminum wires (orange). The overlap of the two \NbSe\ flakes defines the capacitor area (light blue). (d) Side-view schematic of the vdW capacitor and aluminum electrodes (not to scale). (e) Extracted dielectric loss tangents, plotted as $1/\tan{\delta}$. Solid green bars correspond to $\tan{\delta}$ extracted from mean $T_1$ times, except for Device~4 which is the best-$T_1$-derived value. Striped green bars are values extracted from resonator measurements.}
\label{fig:Dielectric}
\end{figure*}

\subsection*{Loss Tangent Extraction}
Having established quantum coherence, we turn to identifying the dominant relaxation mechanism in our fully crystalline METs. In the case that qubit relaxation is limited by dielectric loss, the effective relaxation rate is approximated as \mbox{$1/T_1 \approx 2\pi f_{01}\left[p\tan{\delta} + \left(1-p\right)\tan{\delta_\text{hBN}}\right] = 2\pi f_{01}/Q_\text{qubit}$}, where $Q_\text{qubit}$ is the qubit quality factor. In the MET geometry, the Josephson junction --- characterized by a loss tangent $\tan{\delta}$ --- constitutes the dominant loss channel, while the hBN coupling dielectric, with \mbox{$\tan{\delta_\text{hBN}} = 5.0\times10^{-6}$} taken from Ref.~\cite{wang_hexagonal_2022}, provides an additional contribution. Using the measured energy relaxation times, we evaluate the loss tangents of the vdW Josephson junctions using the relation given above. The extracted values are listed in Table \ref{tab:loss_tangent}.

To independently verify whether the loss originates from dielectric dissipation within the \NbSe/\WSe\ junction rather than from mechanisms such as quasiparticle tunneling, we perform microwave measurements of \NbSe/\WSe\ capacitors at $T = 15~\text{mK}$. These capacitors share the same layer structure as the vdW METs, except that we increase the \WSe\ barrier thickness to $20$--$30~\text{nm}$ to suppress Josephson coupling between the \NbSe\ electrodes and ensure a purely capacitive reactance. We measure $S_{21}$ through a feedline capacitively coupled to either open-ended $\lambda/2$ CPW notch resonators or ones terminated with a vdW capacitor (equivalent circuit shown schematically in Fig.~\ref{fig:Dielectric}(a)). This technique provides a highly sensitive probe of reactances and loss tangents of arbitrary loads appended to a CPW resonator~\cite{kreidel_measuring_2024, chu_measuring_2025_arx}. Here, a capacitive termination modifies the resonator boundary conditions via its complex impedance, resulting in a shift of the resonant frequency from $f_\text{open}$ to $f_\text{term}$, and a change in the internal quality factor from $Q_\text{open}$ to $Q_\text{term}$, as illustrated in Fig.~\ref{fig:Dielectric}(b). By comparing $f_\text{term}$ and $Q_\text{term}$ of a capacitor-terminated resonator to $f_\text{open}$ and $Q_\text{open}$ of a reference resonator fabricated on the same chip, we extract the participation ratio~\footnote{We use the symbol $p$ interchangeably to denote the vdW junction participation ratio in Devices~1--4 and the vdW capacitor participation ratio in Devices~5--8, although these quantities are defined differently.} and loss tangent of each vdW capacitor (see Appendix~\ref{app:dielectric} for further experimental details and definition of participation ratio for capacitor devices). Figures~\ref{fig:Dielectric}(c) and (d) show an optical micrograph and schematic cross section, respectively, of a vdW capacitor integrated with a microwave resonator. In this configuration, the aluminum wires make superconducting contact to the \NbSe\ electrodes, rather than capacitive coupling as in the METs, ensuring negligible additional dissipation or reactance at the termination~\cite{antony_making_2022}.

Figure~\ref{fig:Dielectric}(e) summarizes the extracted loss tangents, plotted as $1/\tan{\delta}$, obtained from mean qubit $T_1$ values (except for Device~4, for which only the best $T_1$ was measured) and from independent capacitor measurements. These values, along with relevant device parameters, are listed in Table~\ref{tab:loss_tangent}. The loss tangents from both measurement types exhibit a comparable spread, indicating consistent device-to-device variation. The absence of trends with device area, \WSe\ thickness or volume, $p$, or frequency suggests that this spread likely arises from extrinsic contamination during fabrication (see Appendix~\ref{app:methods}). Notably, the lowest qubit-derived loss tangent, $1.69\times10^{-5}$ (Device~2), is comparable to --- and even surpasses --- the best resonator-derived value of $3.31\times10^{-5}$ (Device~5). 
Because the resonator measurements directly probe dielectric loss, this agreement supports dielectric loss in the \NbSe/\WSe\ Josephson junction as the dominant mechanism governing energy relaxation at low temperatures in these vdW qubits, rather than quasiparticle tunneling or other loss channels. These results validate our assumption of a dielectric-loss-limited relaxation rate and demonstrate that both qubit- and resonator-based approaches provide consistent metrics for assessing material performance. Further device statistics will be necessary to determine whether the measured loss tangents reflect intrinsic material limits or extrinsic contamination within or surrounding the vdW stack.


\begin{table}[b]
\caption{\label{tab:loss_tangent}Device parameters and extracted loss tangents.}
\begin{ruledtabular}
\begin{tabular}{ccccccc}
 & Area & Thickness & Volume & $p$ & $f$ & $\tan\delta$ \\
Device & ({\textmu}m$^2$) & (nm) & ({\textmu}m$^3$) & (\%) & (GHz) & ($10^{-5}$) \\
\hline
\textbf{Qubits}\\
1 & 20  & 11 & 0.22 & 97 & 4.76 & $3.09\pm0.34$ \\
2 & 8   & 11 & 0.09 & 96 & 5.79 & $1.69\pm0.28$ \\
3 & 24  & 11 & 0.26 & 91 & 5.29 & $12.20\pm2.27$ \\
4 & 20  & 11 & 0.22 & 98 & 5.15 & 15.76$^*$ \\
\hline
\textbf{Capacitors}\\
5 & 200 & 30 & 6.00  & 16 & 5.88 & $3.31\pm0.61$ \\
6 & 303 & 24 & 7.27  & 18 & 4.39 & $5.60\pm0.17$ \\
7 & 309 & 28 & 8.65  & 18 & 6.83 & $9.08\pm1.48$ \\
8 & 65  & 33 & 2.15  & 12 & 6.35 & $14.02\pm3.85$ \\
\end{tabular}
\end{ruledtabular}
$^*$ Indicates value taken from best $T_1$ time.
\end{table}

\subsection*{Temperature Dependence}
\begin{figure}[t]
\centering
\includegraphics[width=0.666666666666\columnwidth]{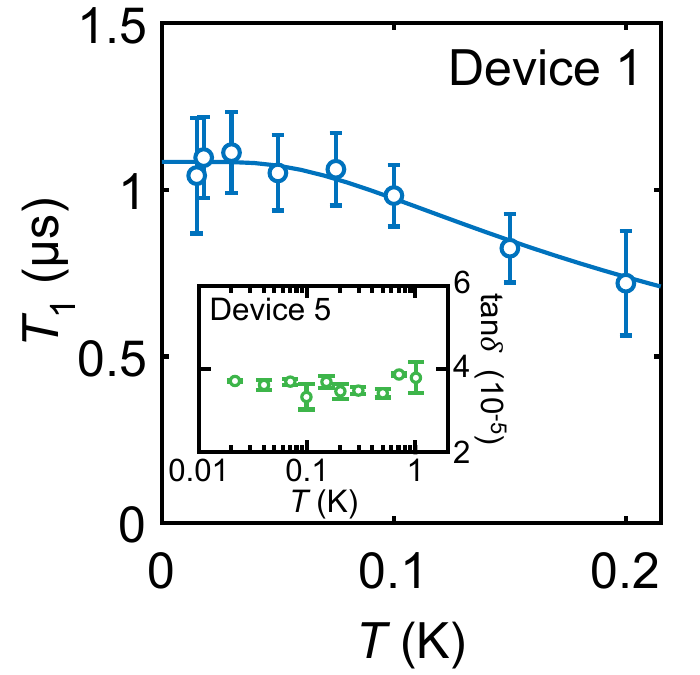}
\caption{Temperature dependence of $T_1$ for Device~1. Mean $T_1$ vs.~$T$ for Device~1. Error bars correspond to standard deviations of distributions taken at each temperature. The solid blue line is a fit to the spin-boson model~\cite{leggett_dynamics_1987}. Inset, dielectric loss vs.~$T$ extracted from capacitor measurements of Device~5.}
\label{fig:Temperature}
\end{figure}

Having established dielectric loss as the dominant relaxation mechanism in our crystalline METs at base temperature, we next examine the temperature dependence of energy relaxation to determine whether this loss channel persists or if additional mechanisms emerge. To this end, we measure $T_1$ distributions of Device~1 as a function of temperature $T$, as shown in Fig.~\ref{fig:Temperature}. We find that the mean relaxation times follow \mbox{$1/T_1 = \tfrac{2\pi E_J}{hQ_\text{qubit}}{|\langle0|\hat{\phi}|1\rangle|}^2 \left[1+\coth{\left(hf_{01}/2k_BT\right)}\right]$} in the measured temperature range, consistent with the spin-boson model of a two-level system with $0\rightarrow1$ transition matrix element $\langle0|\hat{\phi}|1\rangle$ coupled to a dissipative environment~\cite{leggett_dynamics_1987}, with $k_B$ the Boltzmann constant. The close adherence of the measured $T_1$ times to this model rules out quasiparticle-limited relaxation, which would exhibit saturated $T_1$ times out to higher temperatures before abruptly dropping at roughly one third of the junction critical temperature~\cite{martinis_energy_2009} (approximately $2.3~\text{K}$ here). The inset in Fig.~\ref{fig:Temperature} shows a plot of $\tan\delta$ as a function of $T$ extracted from measurements of capacitor Device~5. This quantity remains roughly constant up to $1~\text{K}$, implying that the loss tangent of these \NbSe/\WSe\ heterostructures should not affect $T_1$ at elevated temperatures. While investigation of qubit relaxation times at higher temperatures would elucidate the extent to which dielectric loss dominates relaxation in these qubits, measurements were limited to $T\leq 200~\text{mK}$ due to reduced readout fidelity. Nonetheless, these results show promise for \NbSe/\WSe-based qubits operating above dilution refrigerator temperatures.

\section*{Discussion}
Loss tangents in the low $10^{-5}$ range indicate that \NbSe/\WSe\ Josephson junctions are already approaching practical applicability for superconducting circuits and offer potential for improvement through refined device fabrication and cleaning techniques. Beyond merged-element transmons, these junctions could serve in architectures less sensitive to junction participation, such as conventional transmons with external shunt capacitors or Josephson parametric amplifiers. For example, in conventional transmon designs, the large-area shunt capacitor typically dominates the total qubit capacitance by more than an order of magnitude, so that the Josephson junction contributes only a percent-level fraction of $C_\text{total}$. A $6~\text{GHz}$ conventional transmon incorporating an \NbSe/\WSe\ junction with $p = 1\%$ would exhibit $T_1 = \left(2\pi f_{01} p \tan{\delta}\right)^{-1} \approx 160~\text{\textmu s}$, assuming negligible loss elsewhere in the circuit. In addition, the relatively high critical temperature of \NbSe/\WSe\ junctions can support higher-frequency and higher-temperature operation and even extend the dynamic range of transmon-based thermometry~\cite{lvov_thermometry_2025}. Given the consistency of the $T_1$ temperature dependence with the spin-boson model --- in which the argument $hf_{01}/2k_B T$ determines the extent of the low-temperature $T_1$ plateau --- a $25~\text{GHz}$ \NbSe/\WSe\ qubit would retain 95\% of its $T_1$ at $400~\text{mK}$ and 50\% up to $\approx 1.7~\text{K}$. Because the qubit frequency can be tuned across orders of magnitude through \WSe\ thickness engineering~\cite{balgley_crystalline_2025}, future studies of higher-frequency vdW qubits may enable operation well above millikelvin temperatures and clarify the intrinsic frequency dependence of loss, guiding targeted improvements in fabrication and materials quality. 

The material versatility and potential for higher-temperature operation position vdW qubits as a compelling route towards scalable superconducting quantum technology. Realizing wafer-scale integration will depend on continued progress in materials processing. Current vdW device fabrication relies on the stochastic process of mechanical exfoliation, which randomly produces flakes of varying area and thickness. Recent advances in gold-assisted exfoliation may enable deterministic exfoliation of vdW crystals with controlled area and thickness in wafer-scale arrays~\cite{li_residue-free_2025, olsen_macroscopic_2025}. While the widely practiced ``top-down’’ layer-by-layer fabrication of vdW devices is currently best suited for producing individual devices, recent efforts have demonstrated that this process can be automated~\cite{masubuchi_autonomous_2018, mannix_robotic_2022}. This can allow for rapid testing of different vdW materials combinations, including those with ideal band alignment. Conversely, ``bottom-up’’ growth of vdW materials via increasingly advanced chemical vapor deposition~\cite{zhou_stack_2023} and emerging techniques such as hypotaxy~\cite{moon_hypotaxy_2025} already shows promise for wafer-scale production of vdW flakes with arbitrary thickness and even pre-formed heterostructures such as Josephson junctions. These developments outline a clear trajectory from proof-of-concept vdW qubits to a scalable platform for hybrid and high-temperature superconducting quantum technologies.

\section*{Conclusion}
We demonstrate quantum coherence in first-generation, ultra-compact merged-element transmons made entirely from van der Waals Josephson junctions and identify dielectric loss as the dominant mechanism limiting relaxation up to hundreds of millikelvin in these fully crystalline qubits. Our results highlight the advantages of expanding the materials toolbox for superconducting qubits to include vdW heterostructures, including prospects for higher-temperature operation and pathways toward scalable fabrication. Together, this work establishes van der Waals materials as a promising platform for compact, versatile superconducting quantum devices.

\begin{acknowledgments}
This work was supported by the Laboratory for Physical Sciences and Army Research Office under Contract No.~W911NF-22-C-0021 (NextNEQST SuperVan-2). Development of heterostructure assembly techniques at Columbia was supported by the National Science Foundation through the Columbia University, Columbia Nano Initiative, and the Materials Research Science and Engineering Center (DMR-2011738). J.B.~acknowledges support from the Army Research Office under Grant Number W911NF-24-1-0133. J.P.~acknowledges support from the National Research Foundation of Korea (NRF) grant funded by the Korean government (MSIT) (RS-2024-00358841) and the education and training program of the Quantum Information Research Support Center, funded through the NRF by the MSIT of the Korean government (No.~2021M3H3A1036573). K.W.~and T.T.~acknowledge support from the JSPS KAKENHI (Grants No.~21H05233 and No.~23H02052), the CREST (JPMJCR24A5), JST and World Premier International Research Center Initiative (WPI), MEXT, Japan. J.L.~and A.K.~acknowledge support from the Defense Advanced Research Projects Agency (DARPA) Synthetic Quantum Nanostructures (SynQuaNon) program under Grant Agreement No.~HR00112420343. J.H.~acknowledges support from the Gordon and Betty Moore Foundation’s EPiQS Initiative, Grant No.~GBMF10277. The traveling-wave parametric amplifier (TWPA) used in the qubit experiments was provided by MIT Lincoln Laboratory. The views and conclusions contained in this document are those of the authors and should not be interpreted as representing the official policies, either expressed or implied, of the Army Research Office or the U.S.~Government. The U.S.~Government is authorized to reproduce and distribute reprints for Government purposes notwithstanding any copyright notation herein.
\end{acknowledgments}
 
\appendix

\section{Methods}\label{app:methods}
\subsection{Crystal growth and flake characterization}
WSe$_2$ crystals are grown using a two-step flux synthesis method~\cite{liu_two-step_2023}. Flakes of hBN and WSe$_2$ are mechanically exfoliated from bulk crystals in air and inspected by atomic force microscopy (AFM) to assess surface cleanliness and determine their thickness. While AFM is a sensitive and widely used probe of vdW materials, the presence of physisorbed organic molecules or water, trapped air gaps, or instrumental offsets can obscure the true flake height, introducing a typical uncertainty of around $\pm1$ atomic layer~\cite{kenaz_thickness_2023}. In future work, optical techniques such as second-harmonic generation may be employed to more accurately determine the layer number~\cite{li_probing_2013}.

\subsection{Device assembly and fabrication}
NbSe$_2$ exfoliation and vdW device stacking are performed in a glovebox under a N$_2$-rich atmosphere ($<0.5$~ppm O$_2$ and H$_2$O) to minimize oxidation. Devices are assembled layer-by-layer using polymer-assisted pickup~\cite{wang_one-dimensional_2013}. During vdW heterostructure assembly, polymers used in the stacking process can intercalate between critical interfaces, forming interlayer bubbles that reduce the effective junction area, introduce uncertainty in the Josephson junction geometry, or leave lossy residues that couple to electromagnetic fields and promote qubit relaxation. This likely contributes to the wide spread of extracted loss tangents in the \NbSe/\WSe\ devices, which span nearly an order of magnitude and show no clear correlation with frequency, participation ratio, or geometrical parameters such as \WSe\ area, thickness, or volume (Table~\ref{tab:loss_tangent}). These polymer-related effects may be mitigated by adopting polymer-free vdW stacking techniques~\cite{wang_clean_2023}.

MET and capacitor devices are fabricated by transferring vdW stacks onto pre-patterned resonator chips. Devices~1 and 4--8 were transferred onto solvent-cleaned chips, while Devices~2 and 3 were transferred onto acid-treated chips. The treatment includes a 10 minute piranha etch process (3:1 $\text{H}_2\text{SO}_4$:$\text{H}_2\text{O}_2$) followed by a 10 minute 10:1 BOE dip.

Chips are fabricated from 4-inch high-resistivity Si wafers coated with 200 nm of sputtered Nb. Prior to Nb deposition, the wafers undergo solvent cleaning followed by Piranha etch and BOE treatments. Patterns are written and developed using a standard maskless photolithography process. Circuit structures are subractively defined using inductively coupled plasma (ICP) etching using a CF$_4$/O$_2$/Ar chemistry, which provides high anisotropy and prevents lateral over-etching beneath the resist.

For the MET architecture, each chip contains a feedline coupled to three hanger-type CPW resonators with distinct resonant frequencies, enabling readout of three MET devices per chip. At the end of each resonator, a $100\times100~\text{{\textmu}m}$ landing site is etched out of the ground plane to host the vdW stack, together with two transmission lines used for qubit drive or flux tuning.
The capacitor chips, in contrast, consist of a feedline couple to eight hanger-type CPW resonators. These eight resonators comprise four sets of identical-resonator pairs, with one in each pair serving as the device resonator and the other as a reference. Each pair is designed with a unique resonant frequency. The resonators all have an etched landing site for vdW devices as in the MET chip. On both chips, periodic perforations are etched into the ground plane to serve as vortex traps.

An MMA/PMMA bilayer resist is used for electron-beam lithography to form conections between the Nb circuitry and different layers of the vdW stacks. Before metal deposition, an \emph{in situ} argon ion mill removes surface oxides from the niobium circuitry to ensure good electrical contact. We then deposit a 3-nm-thick Ti adhesion layer and 40--60 nm Al leads by electron-beam evaporation in the same chamber.

For Devices~1--4 (qubits), aluminum leads were deposited directly on the capping hBN layer. Because the hBN in these devices is approximately $40~\text{nm}$ thick, the milling does not fully etch through it, leaving a residual dielectric layer that provides capacitive coupling to the \NbSe\ junction electrodes.
Additional extrinsic loss may arise from surface disorder introduced during ion milling of the hBN coupling regions before metal deposition. Although these regions exhibit significantly smaller energy participation than the junction and thus contribute minimally to total loss, their effect could be further reduced by implementing a two-step milling and deposition process that preserves pristine dielectric interfaces. For Devices~5--8 (capacitors), CHF$_3$/O$_2$ reactive ion etching was used to fully remove the encapsulating hBN and expose the \NbSe\ capacitor electrodes, followed by metal deposition as described above. \emph{In situ} argon ion milling prior to deposition ensures high contact transparency to \NbSe~\cite{antony_making_2022}.

Device areas can be modified with sub-100-nm precision by electron-beam lithography and subsequent etching. For example, Device~4 began with a $23~\text{\textmu m}^2$ junction in which a $0.5~\text{\textmu m}$-wide slit was patterned through the center to divide it into two equal-area junctions. A reactive ion etch using a CF$_4$/Ar gas mixture was employed to etch vertically through the entire stack. Because the etch is predominantly anisotropic, the timing is not critical so long as it results in the stack being fully etched.

\begin{figure*}[t]
\centering
\includegraphics[width=2\columnwidth]{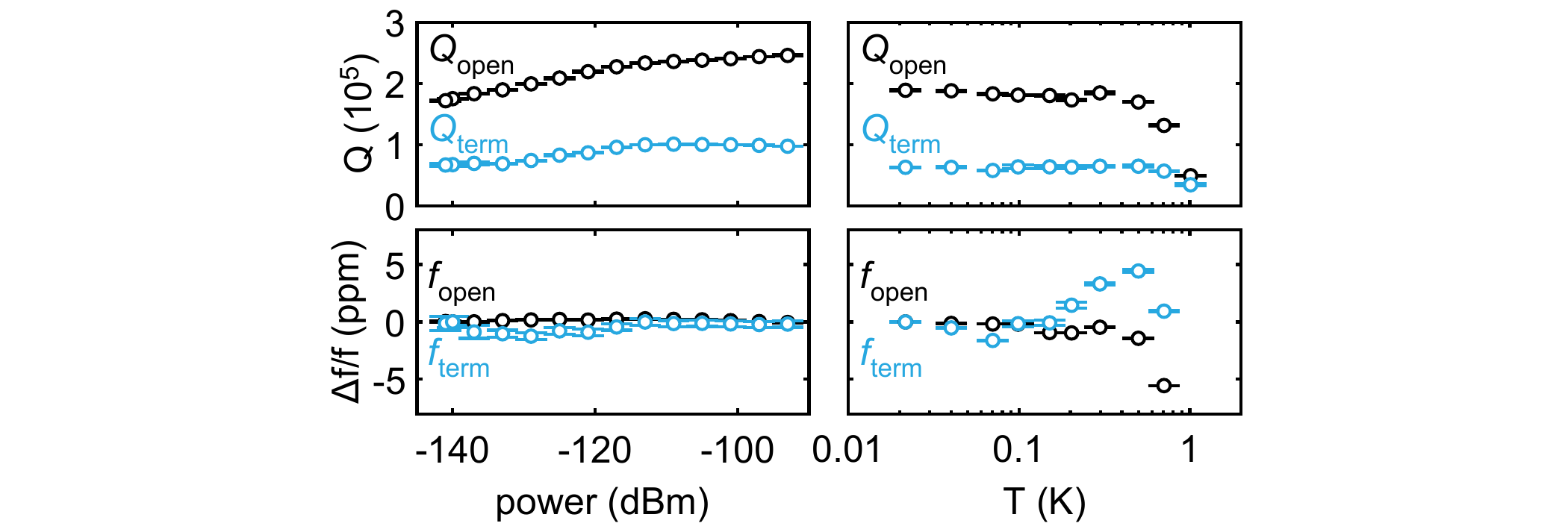}
\caption{Internal quality factors (top row) and relative resonance shifts (bottom row) of a reference resonator ($Q_\text{open}$, $f_\text{open}$) and a capacitor-terminated resonator ($Q_\text{term}$, $f_\text{term}$) as a function of microwave power (left column) and $T$ (right column) for Device~5. The relative resonance shifts here are defined as $\Delta f/f = (f_i - f_{i,0})/f_{i,0}$, where $i = \{\text{open, term}\}$ and the subscript 0 denotes the resonant frequency at the lowest measurement power (left column) or temperature (right column).}
\label{fig:Resonator}
\end{figure*}

\subsection{Cryogenic measurement setup}
Devices are mounted on the cold finger of a dilution refrigerator (Bluefors BF-LD400) with a base temperature below 20 mK. To minimize decoherence from magnetic noise and trapped magnetic flux, the cold finger is enclosed in Cryoperm magnetic shielding. On the input side, the total line attenuation ranges from $70$ to $84~\text{dB}$ depending on the resonance frequency, with a $40~\text{dB}$ attenuator mounted on the mixing chamber to suppress thermal radiation. On the output side, multiple microwave isolators and filters block noise from external sources from the device input. The output signal is amplified using a traveling-wave parametric amplifier (TWPA) from MIT Lincoln Laboratories, mounted to the refrigerator's mixing chamber plate and pumped through a directional coupler. The output signal is subsequently amplified by cascaded low-temperature and room-temperature amplifiers.

For Device~4, the flux-tunable MET, a fixed DC current is applied to an on-chip wire adjacent to the MET to generate magnetic flux, biasing the qubit at its first-order flux-insensitive operating point. Temperature control is achieved using a PID feedback loop implemented through an AC resistance bridge (Lakeshore Model~372) and a $50~\Omega$ heater mounted on the mixing-chamber plate. All measurements are performed only after the temperature has been stabilized for at least 20 minutes.

Qubit lifetime measurements are performed using the pulse sequences depicted in the insets of Figs.~\ref{fig:Coherence}(c)--(e). $\pi$ pulse lengths and amplitudes are determined from Rabi oscillations. $\pi/2$ pulses are defined with the same lengths as the $\pi$ pulses but with half the amplitude. To obtain qubit lifetime statistics, we perform 100 repetitions of interleaved population-inversion, Ramsey, and Hahn-echo sequences. Each sequence in each iteration is taken with 6000 averages for improved signal-to noise. The total 100-iteration experiment takes approximately seven hours. We fit Gaussian lineshapes to the lifetime distributions, whose means and standard deviations are represented by the mean lifetimes and their uncertainties in Table~\ref{tab:qubit_properties}.

\section{Microwave measurements of van der Waals capacitors}\label{app:dielectric}
Here we describe the analytic model used to extract the loss tangent of a capacitive load terminating a CPW resonator, following Ref.~\cite{chu_measuring_2025_arx}. An open-ended CPW of length $\ell$ and characteristic impedance $Z_0$ forms a notch-type $\lambda/2$ resonator with resonant frequency $f_\text{open}$ and internal quality factor $Q_\text{open}$ when capacitively coupled to a CPW feedline.  Here, $f_\text{open} = c\lambda/\sqrt{n_\text{eff}}$, where $\lambda = 2\ell$ is the resonant wavelength of the resonator, $n_\text{eff}$ is the effective dielectric constant of the environment, and $c$ is the speed of light in vacuum. The terminating capacitive load has complex impedance $Z_L = r + jX$, where $r$ and $X$ denote its resistive and reactive components, respectively.

The fundamental resonant frequency of the capacitively terminated resonator $f_\text{term}$ is obtained by solving the transmission-line equations~\cite{pozar_microwave_2012} under the boundary conditions described in Ref.~\cite{chu_measuring_2025_arx}, yielding
\begin{align}
f_\text{term} = \left(\frac{1}{\pi}\tan^{-1}\frac{Z_0}{X}\right)f_\text{open}. \label{eq:fr}
\end{align}
Integrating the electric and magnetic energy distributions of the loaded resonator gives the energy participation ratio,
\begin{align}
p = \frac{\lvert \sin \phi \rvert}{\phi + \lvert \sin \phi \rvert}, \label{eq:p}
\end{align}
where $\phi = 2\pi f_\text{term} / f_\text{open}$ represents the phase shift between the incident and reflected waves at the load termination. The corresponding expression for the internal quality factor is
\begin{equation}
Q_\text{term}^{-1} = 2p\tan\delta + (1-p)Q_\text{open}^{-1}. \label{eq:Qi_p}
\end{equation}

To extract $\tan{\delta}$, we first measure the complex transmission coefficient $S_{21}$ of the feedline at frequencies around the designed value of $f_\text{open}$ for an unloaded reference resonator fabricated identically to, and on the same chip as, the capacitively terminated resonator. Using a ``circle fitting'' procedure~\cite{probst_efficient_2015}, we obtain experimental values of $f_\text{open}$ and $Q_\text{open}$. We then repeat this measurement for the loaded resonance to obtain $f_\text{term}$ and $Q_\text{term}$, with which we can solve Eqns.~(\ref{eq:p}) and (\ref{eq:Qi_p}) for $\tan{\delta}$. The error associated with the goodness of fits from circle-fitting the reference and device resonances is propagated accordingly to provide error bars for $\tan{\delta}$, as plotted in Fig.~\ref{fig:Dielectric}(e). We perform this measurement at various microwave powers and temperatures, as exemplified in Fig.~\ref{fig:Resonator} for Device~5. We see that the quality factors increase with power commensurate with saturation of two-level systems in the measurement circuit, while they decrease with increasing temperature due to increasing quasiparticle density. Meanwhile, the relative frequency shifts (see caption) are only a few parts per million for these powers and temperatures. Loss tangent values in Fig.~\ref{fig:Dielectric}(e) and inset in Fig.~\ref{fig:Temperature} are extracted from internal quality factors measured at single-photon-level powers, corresponding to approximately $-145~\text{dBm}$.

\begin{figure}[tb]
\centering
\includegraphics[width=1\columnwidth]{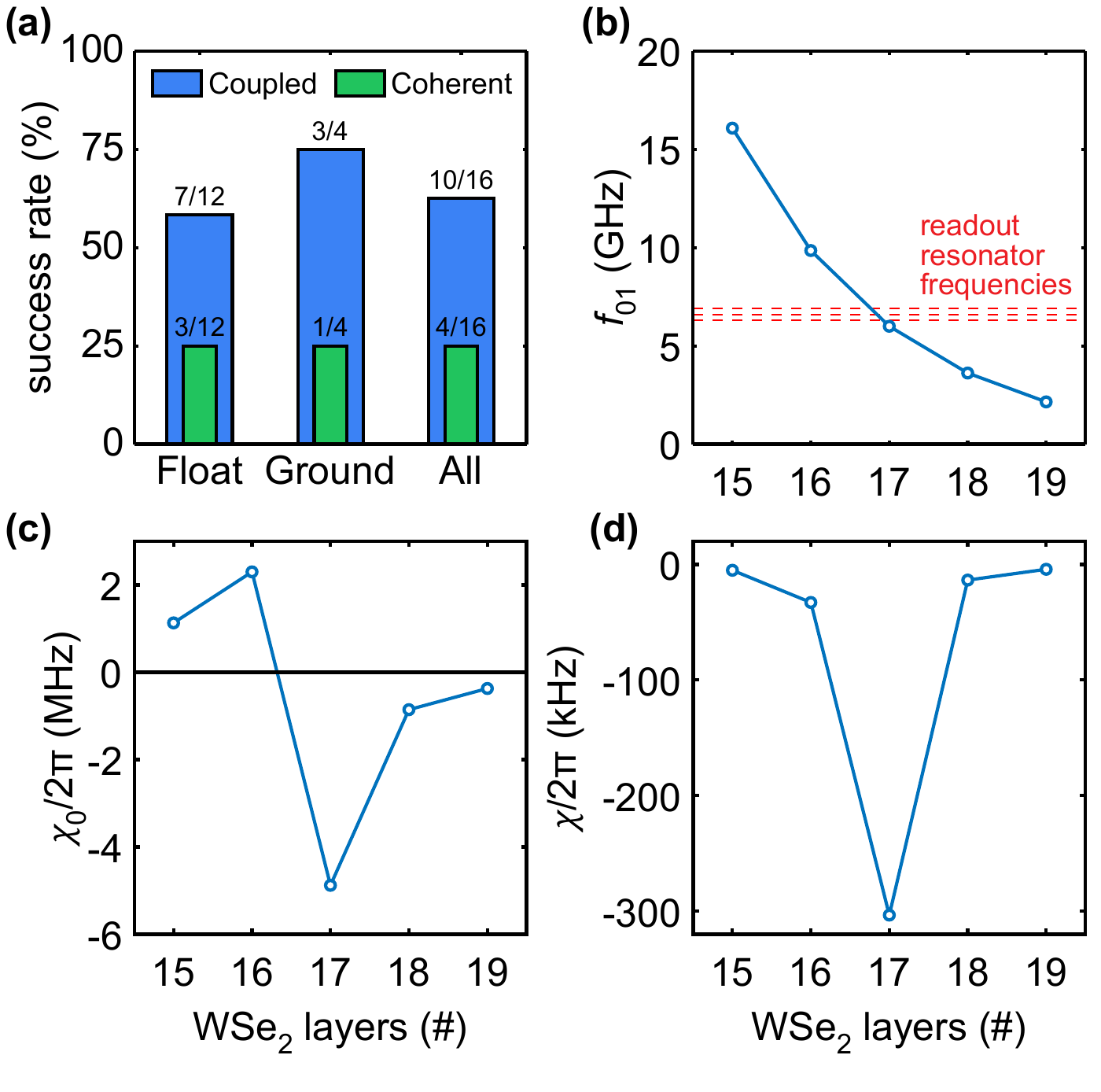}
\caption{Success rates of measuring vdW qubits. (a) Percentage of all measured vdW qubits successfully showing qubit-resonator coupling (blue) and coherence (green). (b)--(d) Calculated qubit frequencies, Lamb shift, and dispersive shift, respectively, for METs with 15--19 layers of \WSe. Red lines in (b) indicate the designed readout resonator frequencies. The black horizontal line (c) marks $\chi_0 = 0$ for clarity.}
\label{fig:Success}
\end{figure}

\begin{table}[b]
\centering
\begin{tabular}{|c|c|c|c|c|}
\hline
Device & 1 & 2 & 3 & 4 \\ \hline
$T_1$ ({\textmu}s) & 1.11 & 1.67 & 0.27 & 0.20$^*$ \\ \hline
Acid-treated & \textcolor{red}{$\times$} & \checkmark & \checkmark & \textcolor{red}{$\times$} \\ \hline
$E_J/E_C > 30$ & \checkmark & \textcolor{red}{$\times$} & \checkmark & \checkmark \\ \hline
Floating & \checkmark & \checkmark & \textcolor{red}{$\times$} & \checkmark \\ \hline
Fixed-frequency & \checkmark & \checkmark & \checkmark & \textcolor{red}{$\times$} \\ \hline
\end{tabular}
\caption{Qubit design parameters. Checkmarks indicate that the criterion is satisfied, while red $\times$ indicate the opposite. All $T_1$ times are mean distribution values except $*$ which is the best measured time.}
\label{tab:design}
\end{table}







\section{Success rates of measuring vdW qubits}\label{app:success}

Devices~1--4 represent a subset of all vdW METs prepared for this study. In total, sixteen devices were fabricated, twelve of which were floating and four grounded. Of these, ten exhibited qubit-resonator coupling, as evidenced by a measurable Lamb shift (also known as ``punchout'') of the resonator with increasing readout power, and four of those showed coherence. Additionally, four of the twelve floating qubits were etched using reactive ion etching to create SQUID loops or modify the junction area, and only one of these (Device~4) displayed coherence. A summary of the fabrication yield, expressed as the percentage of qubits exhibiting coupling and coherence, is shown in Fig.~\ref{fig:Success}(a), while the different device design parameters are tabulated in Table \ref{tab:design}.

While fabrication defects can suppress coherence, we attribute the $62.5\%$ rate of observing qubit-resonator coupling and the $25\%$ rate of coherence primarily to uncertainty in the \WSe\ thickness, which directly impacts the qubit-resonator detuning and dispersive shift (see Appendix A).

Each qubit chip contains three readout resonators with $f_{RO} = 6.3~\text{GHz}$, $6.6~\text{GHz}$, and $6.9~\text{GHz}$, respectively. Based on measured trends in critical current density versus \WSe\ thickness~\cite{balgley_crystalline_2025}, a 17-layer-thick \WSe\ junction is expected to yield $f_{01} \approx 6.0~\text{GHz}$, giving $\Delta/2\pi$ on the order of a few hundred MHz (Fig.~\ref{fig:Success}(b)). This configuration allows engineered Lamb shift on the order of a few MHz and a dispersive shift of roughly half the resonator linewidth ($\chi/2\pi\approx300~\text{kHz}$) with reasonable coupling capacitances of a few femtofarads. However, if the \WSe\ thickness deviates by even one layer --- for example, to 16 or 18 layers --- $\Delta/2\pi$ can increase by nearly an order of magnitude, dramatically reducing both $\chi_0$ and $\chi$ (Figs.~\ref{fig:Success}(c) and (d)). Numerically, this can yield dispersive shifts of only tens of kHz, lowering readout fidelity below the system noise floor. As discussed in Appendix \ref{app:methods}, this uncertainty can be eliminated with improved vdW layer-number metrology.


\bibliography{arXiv_v1}

\end{document}